\documentclass[12pt]{article}
\begin{document}

\title{AdS/CFT   Correspondence and string/gauge duality }

\author { Henrique Boschi-Filho and Nelson R. F. Braga \\
 Instituto de F\'\i sica, Universidade Federal  do Rio de Janeiro\\
\it Caixa Postal 68528, 21941-972  Rio de Janeiro, RJ, Brazil 
\\E-mails: boschi@if.ufrj.br and braga@if.ufrj.br  }


\maketitle

\abstract{The AdS/CFT   correspondence is an exact duality between string 
theory in anti-de Sitter space and conformal field theories on its 
boundary.  
Inspired in this correspondence some relations between strings
and non conformal field theories have been found. 
Exact dualities in the non conformal case are intricate
but approximations can reproduce important physical results. 
A simple approximation consists in taking
just a slice of the AdS space with a size related to an energy scale.
Here we will discuss how this approach can be used to reproduce the scaling
of high energy QCD scattering amplitudes.
Also we show that very simple estimates for glueball mass ratios can emerge from 
such an approximation. }

\section{Introduction}

The AdS/CFT   correspondence is a duality between the large $N$ limit of $SU(N)$
superconformal field theories and string theory in a higher 
dimensional anti-de Sitter spacetime\cite{Malda}.   
In this correspondence the AdS space shows up both as a near horizon 
geometry of a set of coincident D-branes or as a solution of supergravity. 
Precise prescriptions for the realization of the AdS/CFT   
correspondence were formulated\cite{GKP,Wi} by considering  Poincare 
patches of AdS space. The Poincare coordinate system allows
a simple definition for the boundary where the conformal field theory 
is defined. In this correspondence, bulk fields act as classical sources for boundary 
correlation functions.\cite{Pe}$-$\cite{FD}
One of the striking features\cite{Wi} of this  correspondence is that it 
is a realization of the holographic principle: 
``The degrees of freedom of  a quantum theory with gravity can be mapped 
on the corresponding boundary".\cite{HOL1}$-$\cite{BS}
 
String theory, presently the main candidate to describe the fundamental 
interactions in a unified way, was originally proposed as a model for strong 
interactions with the successful results of Veneziano amplitude in the Regge 
regime\cite{V}$-$\cite{SH}. 
However this model was unable to reproduce the experimental 
hard and deep inelastic scatterings. 
These behaviours were latter explained by QCD which is today the theory
accepted to describe strong interactions.
Despite of its success in high energies, QCD is non perturbative in 
the low energy regime, where one needs involved lattice calculations. 
In fact QCD and strings may hopefully be viewed as complementary theories.
A strong indication found by 't Hooft\cite{Planar} is the relation between 
large $N$ $SU(N)$ gauge theories and strings\cite{Pol}.
A recent remarkable result in the direction of understanding this
duality was the AdS/CFT   correspondence found by Maldacena.

One of the long standing puzzles for the string description of strong interactions
is the high energy scattering at fixed angles.
For string theory in flat space such a process is soft (the amplitudes decay 
exponentially with energy\cite{GSW} while both experimental 
data and QCD theoretical predictions\cite{QCD1,BRO} indicate a hard behavior 
(amplitudes decaying with a power of energy).
A solution for this puzzle was  proposed recently by 
Polchinski and Strassler\cite{PS}, inspired in the AdS/CFT   correspondence,
considering strings not in flat space but rather 
in a 10 dimensional space which assimptotically tends to an AdS$_5$ times a compact 
space. 
This space is considered as an approximation for the space 
dual to a confining gauge theory which can be associated with QCD.
An energy scale was introduced and identified with the lightest 
glueball mass. 
Then they found the glueball high energy scattering amplitude 
at fixed angles (with the correct QCD scaling) integrating the string amplitude 
over the warped AdS extra dimension weighted by the dilaton wave function.
Other approaches to this problem have also been discussed 
recently\cite{GI}$-$\cite{AN}. 
Important related
discussions on deep inelastic scattering of hadrons\cite{PS2}
and on exclusive processes in QCD\cite{Brod}, 
both inspired in the AdS/CFT correspondence appeared recently.

In this paper we are going to review recent results 
regarding an approximate relation between strings and QCD.
In particular we find the scaling of high energy glueball scattering 
at fixed angles and their masses. 
This paper is organized in the following way. In section 2 we will
review  the relation between AdS space and D-brane spaces coming from string
theory. In section 3 we first present a mapping betwen AdS bulk and boundary states
for scalar fields. Then we show in section 4 how this mapping can be used to find
the same high energy scaling as that of glueball scattering, from the bulk theory.
In section 5 we use this mapping to obtain an estimate for the ratio
of scalar glueball masses.

\section{Branes and AdS space}
 
An  $\,n+1$ dimensional AdS space can be defined 
as the hyperboloid ($R\,=\,$constant)

\begin{equation}
X_0^2 + X_{n+1}^2 - \sum_{i=1}^n X_i^2\,=\,R^2
\end{equation}

\noindent in a flat $n+2$ dimensional space with measure
\begin{equation}
ds^2\,=\, - d X_0^2 - dX_{n+1}^2 + \sum_{i=1}^n dX_i^2.
\end{equation}

Usually the AdS space is represented by the so called global 
coordinates $\,\rho,\tau,\Omega_i\,$ 
given by \cite{Pe,Malda2}

\begin{eqnarray}
\label{global}
X_0 &=& R \,\sec\rho\, \cos \tau \nonumber\\
X_i &=& R \,\tan \rho\, \,\Omega_i\,\,\,;\qquad\,\,\,\,\,\,\,
(\,\sum_{i=1}^n \,\Omega^2_i\,=\,1\,) \nonumber\\
X_{n+1} &=& R \sec \rho \,\sin\tau \,.
\end{eqnarray}

\noindent These coordinates are defined in the ranges 
$0\le \rho <\pi/2$ and $0\le\tau< 2\pi\,$ and  the line element
reads

\begin{equation}
ds^2 \,=\, {R^2 \over cos^2 (\rho )} \Big( -d\tau^2 + d\rho^2 + 
sin^2 (\rho )d\Omega^2 \Big)\,\,.
\end{equation}

\noindent  
Note that the timelike coordinate $\tau $ in the above metric has a finite range.
So in order to identify it with the usual time it is necessary to unwrap it. 
This is done taking copies of the original 
AdS space that together represent the AdS covering space\cite{HE}. 
From now on we take this covering space and call it simply as AdS space.

A consistent  quantum field theory in AdS space\cite{QAdS1,QAdS2} requires the 
addition of a boundary  at spatial infinity: $\rho \,=\, \pi/2\,$.
This compactification of the space makes it possible to impose appropriate 
conditions and find a well defined Cauchy
problem. Otherwise massless particles could go to or come from 
spatial infinity in finite times. 
Possible energy definitions in AdS spaces, relevant for the AdS/CFT correspondence
have also been discussed\cite{MR}.

Poincar\'e coordinates $\,z \,,\,\vec x\,,\,t\,$  are 
more useful for the study  of the AdS/CFT   correspondence. 
These coordinates are defined by

\begin{eqnarray}
\label{Poincare}
X_0 &=& {1\over 2z}\,\Big( \,z^2\,+\,R^2\,
+\,{\vec x}^2\,-\,t^2\,\Big)
\nonumber\\
X_j &=& {R x^j \over z}\,\,\,;\,\,\,\,\,\,\,\,\,\,\,\qquad (j=1,...,n-1)
\nonumber\\
X_n &=& - {1\over 2z}\,
\Big( \,z^2\,-\,R^2\,+\,{\vec x}^2\,-\,t^2\,\Big)
\nonumber\\
X_{n+1} &=& {R t \over z}\,,
\end{eqnarray}

\noindent where $\vec x $ has $n-1$ components and 
 $0 \le  z < \infty $. In this case the $\,AdS_{n+1}\,$ metric reads

\begin{equation}
\label{metric}
ds^2=\frac {R^2 }{( z )^2}\Big( dz^2 \,+(d\vec x)^2\,
- dt^2 \,\Big)\,.
 \end{equation}

The AdS boundary $\rho \,=\, \pi/2$ in global coordinates corresponds 
in Poincar\'e coordinates to the surface $\,z\,=\,0\,$ 
plus a ``point'' at infinity ($z\,\rightarrow\,\infty\,$).
The surface $\,z\,=\,0\,$ defines a Minkowiski space
with coordinates $\vec x$ , $t$. 
The connectedness of the AdS boundary was discussed by Witten and Yau\cite{WY}.

The point at infinity can not be accommodated in the original Poincare 
chart\cite{BB1,BB2} so that  we have to introduce a second coordinate system
to represent it properly. The consistency of the addition of a second chart
requires ending the first chart at a finite $z = z_{max}\,$.
At this surface one needs to impose boundary conditions to guarantee
the analiticity of fields. The imposition of boundary conditions will
be essential to the discussion of the next section.

Finally  it is interesting to see how AdS space 
shows up when we study D-brane systems in string theory.  
The ten dimensional metric generated by  $N$ coincident D3-branes
can be written as\cite{HS,GKP} 
 
\begin{equation}
\label{branemetric}
ds^2 \,=\, \Big( 1 + {R^4\over r^4} \Big)^{-1/2} ( -dt^2 + d{\vec x}^2 ) +  
\Big( 1 + {R^4\over r^4} \Big)^{1/2} (dr^2 + r^2 d\Omega^2_5 )
\end{equation}

\noindent where $R^4 \,=\, N/ 2\pi^2 T_3$ and 
$T_3$ is the tension of a single D3-brane.  Note that there is a horizon at $r = 0$.

Now looking at the near horizon region $r << R $ 
the metric (\ref{branemetric}) can be approximated as:

\begin{equation}
ds^2 \,=\, {r^2 \over R^2} ( -dt^2 + d{\vec x}^2 ) +  
{R^2\over r^2}dr^2 + R^2 d\Omega^2_5 \,.
\end{equation}

\noindent Changing the axial coordinate according to: $ z = R^2/r$
this metric takes the form

\begin{equation}
\label{metric3}
ds^2=\frac {R^2 }{ z^2}\Big( dz^2 \,+(d\vec x)^2\,
- dt^2 \,\Big)\,+\,R^2 d\Omega^2_5.
\end{equation}

\noindent corresponding to a five dimensional AdS space (in Poincar\'e
coordinates) times a five sphere:  AdS$_5\times S^5\,$.

\section{Bulk boundary mapping}
In the  AdS/CFT  framework there is a correspondence between  
(on shell) string theory in the AdS bulk and 
(off shell) conformal field theory on the boundary. 
This represents a realisation of the holographic principle
which  asserts that the degrees of freedom of a quantum theory with 
gravity in some space can be represented on the corresponding boundary.
Based on this principle one can speculate on the possibility of finding a map
between quantum states of theories defined in the bulk and on the boundary
of a given space.
Here we will find a map between AdS bulk and boundary states
considering the simple situation of free scalar fields\cite{BB3}.
We will later identify the bulk scalar field with the dilaton
which is a massles string excitation.

A free scalar field, like the dilaton, in a Poincar\'e AdS$_5$ chart 
$0\,\le\,z\,\le z_{max}\,$ has the form\cite{BB1}
 
\begin{eqnarray}
\label{QF}
\Phi(z,\vec x,t) &=& \sum_{p=1}^\infty \,
\int { d^3 k \over (2\pi)^{3}}\,
{z^{2} \,J_2 (u_p z ) \over z_{max} w_p(\vec k ) 
\,J_{3} (u_p z_{max} ) }\nonumber\\
&\times& \lbrace { {\bf a}_p(\vec k )\ }
 e^{-iw_p(\vec k ) t +i\vec k \cdot \vec x}\,
\,+\,\,h.c.\rbrace\,,
\end{eqnarray}

\noindent with $w_p(\vec k ) \,=\,\sqrt{ u_p^2\,+\,{\vec k}^2}\,$ and   
$u_p$ defined by 
\begin{equation}
\label{up}
u_p z_{max}\,=\, \chi_{_{2\,,\,p}}\,,
\end{equation}

\noindent where $\chi_{_{2\,,\,p}}$ are the zeros of 
the Bessel function $ J_2 ( y )$. 

The operators 
${\bf a}_p\, ,\;{\bf a}^{\dagger}_p \,$ satisfy the commutation relations
\begin{equation}
\label{canonical1}
\Big[ {\bf a}_p(\vec k )\,,\,{\bf a}^\dagger_{p^\prime}({\vec k}^\prime  )
\Big]\,=\, 2\, (2\pi)^{3} w_p(\vec k )   
\delta_{p\,  p^\prime}\,\delta^{3} (\vec k -
{\vec k}^\prime )\,.
\end{equation}

As the coordinate range $z = z_{max}$ is arbitrary we can take it as large as we want
so that most of the AdS space is described by one Poicar\'e chart. This way
we will not need to use the second Poincar\'e chart explicitly.
The size $z = z_{max}$ corresponds to an energy scale as we 
will discuss bellow. From now on we call the Poincar\'e chart toghether with 
boundary conditions at $z = z_{max}$ as the AdS slice.
  
On the four dimensional boundary ($ z = 0)$ of the AdS slice we consider massive 
scalar fields $\Theta(\vec x,t)\,$
whose corresponding creation-annihilation operators are assumed to 
satisfy the algebra  
\begin{equation}
\label{canonical2}
\Big[ {\bf b}( \vec K )\,,\,{\bf b}^\dagger ({ \vec K }^\prime  )
\Big]\,=\, 2 (2\pi)^{3} \, w( \vec K ) \,\delta^3 ( \vec K -
{ \vec K}^\prime ) \,,
\end{equation}

\noindent where $ w(\vec K ) = \sqrt{ {\vec K}^2 + \mu^2}$ and $\mu$ is the mass
of the field $\Theta$.

Note that if the above bulk and boundary field theories 
had continuous momenta it would be impossible
to find a one to one mapping between the corresponding quantum states
since they are defined in different dimensions.
However, as we consider just a slice of AdS, 
the spectrum of the momentum $u_p\,$ associated with the axial 
direction $z\,$ is discrete. 
Then naturally the continuous part of bulk and boundary momenta 
$\vec k$ and $\vec K$ have the same dimensionality. 
So this discretization makes it possible to establish 
a one to one mapping between bulk
and boundary momenta $(\vec k , u_p)$ and $\vec K$. 
Further we assume a trivial mapping between their angular parts
and then look for a relation between  their moduli
$K \equiv \vert \vec K \vert \, , \,k \equiv \vert \vec k \vert \,$.
It is important to mention that the complete one to one bulk boundary
mapping is only possible in the conformal limit $\mu \rightarrow 0$ of
the $\Theta $ field. 
In the discussion bellow we will be interested in an approximate
map between the dilaton and massive fields so we will let
$\mu \ne 0$. Naturally the massless limit is closest to the original 
AdS/CFT correspondence.

We may introduce a sequence of energy scales ${\cal E}_1 \,,\,
{\cal E}_2\,,{\cal E}_3\,...$ 
in order to map each interval of the boundary momentum modulus
${\cal E}_{p-1} < K \le {\cal E}_p \,$ with $p=1,2,...$  
into the entire range of the transverse bulk momentum modulus $ k$,  
corresponding to some fixed axial momentum $u_p$. 
This mapping between bulk and boundary momenta allow us to
map the corresponding creation-annihilation operators 
(\ref{canonical1},\ref{canonical2}).

For the first energy interval, corresponding to $u_1$,  defined as 
$0 \le K \le {\cal E}_1\,$, we can write\cite{BB3} 
\begin{eqnarray}
\label{ab}
k\,{\bf a}_1 ( \vec  k ) 
&=& K \,{\bf b}( \vec K  ) \nonumber\\
k\,{\bf a}^\dagger_1 ( \vec k ) 
&=& K\,{\bf b}^\dagger ( \vec K  )\,.
\end{eqnarray}

\noindent This mapping must preserve the physical consistency
of both theories. In particular Poincare invariance should not be broken
neither for the boundary theory nor for the bulk theory at a fixed $z$.
This is guaranteed imposing that the canonical commutation relations 
(\ref{canonical1},\ref{canonical2}) are preserved by the mapping (\ref{ab}).
Then substituting eq. (\ref{ab}) into relation (\ref{canonical1}) and using 
eq. (\ref{canonical2}) 
we find an equation in terms of the moduli of bulk and boundary
momenta which solution can be written as\cite{BB3}
\begin{equation}
\label{completo}
k = {u_1 \over 2} 
\,\Big[ \,{ {\cal E}_1  +\sqrt{{\cal E}_1^2 + \mu^2 } 
			\over  K + \sqrt{K^2 + \mu^2}}
- { K + \sqrt{K^2 + \mu^2}\over {\cal E}_1  
+\sqrt{{\cal E}_1^2 + \mu^2 }}\,\Big]\,.
\end{equation}

\noindent Similar relations can be obtained for the other intervals 
${\cal E}_{p-1} < K \le {\cal E}_p \,$ with $p=2,3,...$.
This way we found a one to one map between bulk and boundary 
creation-anihilation operators. This allows us to construct a similar
map for  quantum states. This will be used in the next section
in order to relate bulk and boundary scattering amplitudes. 

One might wonder if the trivial mapping $\vec k = \vec K$ would also 
be a solution.
However this would not provide a one to one mapping between the  
entire bulk momenta
$(\vec k , u_p )$ and  
the boundary momenta $\vec K$ for all different values of $p$
as long as there is only one boundary field with mass $\mu$. 

The momentum operators in the bulk and boundary theories are respectively:
\begin{eqnarray} 
(\vec P ,u) &=& \sum_p \int {d^3k \over 2 (2\pi)^3 } 
{{\bf a}^\dagger_p ( \vec k ) {\bf a}_p ( \vec k ) \over \sqrt{k^2 + u_p^2}}
 (\vec k , u_p )\\
\vec \Pi &=&  \int {d^3K \over 2 (2\pi)^3 } 
{{\bf b}^\dagger ( \vec K ) {\bf b} ( \vec K ) \over \sqrt{K^2 + \mu^2}}
 \vec K
\end{eqnarray}
\noindent Note that Poincare invariance in the $\vec x$ directions holds 
both in 
the boundary and bulk theories since the canonical commutation relations 
(\ref{canonical1}) and (\ref{canonical2}) are preserved by the mapping.
 
\section{High energy scalling}

In this section we are going to study the scattering of scalar particles  
at high energy using the map found in the previous section.
This is inspired in the gauge /string duality sugested by the 
AdS/CFT   correspondence, where the scaling of the high energy scattering 
of glueballs was obtained\cite{PS} from string theory using the duality between 
glueballs and dilatons\cite{OZ1}.
Glueballs in QCD correspond to composite operators. So we do not expect the free 
boundary scalar field of the previous section to give a complete 
description of their dynamics. However, we will see that for high energies the map
leads to the same scaling as that of QCD glueballs. This may be an indication that 
for this regime the assymptotic states of the glueballs may be approximated by
free scalar fields.

So we will consider the mapping  of the previous section in the following way: 
on the boundary we will approximate the assimptotic behaviour of the glueball  
operators by free massive scalar fields in four dimensions.
In the bulk we take the massles scalar fields in the AdS slice as representing 
the dilaton.  
A high energy scattering  on the four dimensional AdS boundary
will be mapped into a scattering process of dilaton states 
in the bulk\cite{BB4}.

The equation (\ref{completo}) 
is understood as the relation between bulk and 
boundary momenta for the particles involved in the scatterings.
Identifying $\mu$ with
the mass of the lightest glueball and choosing the AdS size as  
\begin{equation}
z_{max} \,\sim\, {1\over \mu}\,
\end{equation}
 
\noindent we find that $u_1 \sim \mu $, once $z_{max} \,\sim 1/u_1$
according to eq. (\ref{up}). Note that the size $z_{max}$ 
can then be interpreted as
an infrared cutoff for the boundary theory.

Further, we can take  ${\cal E}_1\,$ large enough so that the 
momenta associated with the high energy glueball scattering can 
fit into the region $\mu \ll K \ll {\cal E}_1$. 
Then we can  approximate relation (\ref{completo}) as
\begin{equation}
\label{Kk}
k \,\,\approx\,\, { {\cal E}\, \mu  \over 2 \, K}\,, 
\end{equation}

\noindent where we defined ${\cal E}_1\,\equiv\,{\cal E}\,$ and 
disregarded the other energy intervals
associated with higher axial momenta $u_p\,,\,p\ge\,2\,$.
Note that this mapping together with the conditions 
$\mu \ll K \ll {\cal E} $ imply  that $\mu \ll k \ll {\cal E} $.

Choosing the string scale to  be of the same order of the high 
energy cut off of the boundary theory, i.e., 
${\cal E}\,\sim 1/ \sqrt{\alpha^\prime}$,
we find that the momenta $k$ associated with string theory 
correspond to a low energy
regime well described by supergravity approximation.

The equations (\ref{ab}) and (\ref{Kk})  represent a one to one 
holographic mapping
between bulk dilaton  and boundary glueball states.
Now we are going to use these equations to relate the corresponding 
scattering amplitudes.

Let us consider, in the bulk string theory, a scattering  of
2 particles in the initial state and $m$ particles in the final state,
with all particles having axial  momentum  $u_1$. The $S$ matrix reads 
\begin{eqnarray} 
S_{Bulk} &=&  \langle \, {\vec k}_3\,,u_1;\,...; \,{\vec k}_{m+2}\,,u_1\,
;\,out \vert {\vec k}_1,\,u_1\, 
;\,{\vec k}_2 \,,u_1;in\,\rangle \nonumber\\
&=&  \langle \, 0\,\vert \,{\bf a}_{out} ( {\vec k}_3 )\,... \,{\bf a}_{out}
 ({\vec k}_{m+2}) \,  {\bf a}^+_{in }
( {\vec k}_1) \,{\bf a}^+_{in} ({\vec k}_2) \,\vert \, 0\, \rangle 
\,,\nonumber\\
& &
\end{eqnarray}

\noindent where ${\bf a} \equiv {\bf a}_1$ and the $in$ and $out$ 
states are defined as 
$ \vert \vec k \,,\,u_1\,\rangle 
\,=\, {\bf a}^+ (\vec k ) \vert 0 \rangle \,$.

Now using the mapping between creation-annihilation operators 
(\ref{ab}) one can rewrite the above $S$ matrix in terms of 
boundary operators.
Considering fixed angle scattering, we take the bulk momenta 
to be of the form
$ k_i \,=\, \gamma_i k$ and the boundary momenta 
$ K_i \,=\, \Gamma_i K$, where
$\gamma_i $ and $\Gamma_i$ are constants with $i\,=\,1,2,...,m+2\,$. 
Then

\begin{eqnarray}
S_{Bulk}&\sim &  \langle  0 \vert \,{\bf b}_{out} ( {\vec K}_3 )\,... \,
\,{\bf b}_{out} ({\vec K}_{m+2})   
\,{\bf b}^+_{in }( {\vec K}_1)\, {\bf b}^+_{in} ({\vec K}_2) \vert 0 \rangle  
\,\Big({ K \over k}\Big)^{m + 2 } \nonumber\\
&\sim& \, \langle  \,{\vec K}_3 \,,... \,{\vec K}_{m+2},\,out \,\vert \,{\vec K}_1 \,,
{\vec K}_2 \,,in\,\rangle \, \Big( { K \over k} \Big)^{m+ 2} K^{(m+2)(d-1) }\,,
\end{eqnarray}

\noindent where the composite operators on the boundary have  scaling dimension $d$
and then their $in$ and $out$ states are 
$ \vert \vec K \,\rangle \,\cong\, K^{ 1 - d } {\bf b}^+ (\vec K ) \vert 0 \rangle \,,$
within the regime $K \gg \mu$.

Using the relation (\ref{Kk}) between bulk and boundary momenta we get
\begin{equation}
S_{Bulk} \, \sim \,  
 S_{Bound.} \,\,\Big( {\sqrt{\alpha^\prime} \over \mu }\Big)^{m+2} \,\, K^{(m+2)(1+d)}
\end{equation}

As the scattering amplitudes ${\cal M}$ are related to  the corresponding $S$ matrices 
(for non equal $in$ and $out$ states) by
\begin{eqnarray}
S_{Bulk} &=& {\cal M}_{Bulk} \, \delta^4 (  k_1^\rho +  k_2^\rho -  k_3^\rho - \,...\,- 
 k_{m+2}^\rho )
\nonumber\\
 S_{Bound.} &=& {\cal M}_{Bound.} \, \delta^4 ( K_1^\rho +  K_2^\rho 
- K_3^\rho - ... -K_{m+2}^\rho  )\nonumber\\
& &
\end{eqnarray}

\noindent we find a relation between bulk and boundary scattering amplitudes\cite{BB4}
\begin{eqnarray}
\label{Mb}
{\cal M}_{Bound.} &\sim& {\cal M}_{Bulk}\,\,  S_{Bound.}\,\, (\, S_{Bulk}\,)^{-1} 
\Big( { K\over k} \Big)^4 \nonumber\\
&\sim&  {\cal M}_{Bulk}\, \,K^{8 -  (m+2)(d + 1)  } \,\,\,
\Big( {\sqrt{\alpha^\prime} \over \mu }\Big)^{2 - m}\,\,.
\end{eqnarray}

Now we must evaluate the bulk amplitude from the string low energy 
effective action. At energies much lower than the string scale 
$1/\sqrt{\alpha^\prime}\,$ type IIB string theory can be described  
by  the supergravity action\cite{GKP,Po}
\begin{equation}
S \,=\, {1\over 2 \kappa^2} \int d^{10}x \sqrt{G} e^{-2\Phi} \Big[
{\cal R} + G^{MN} \partial_M \Phi \partial_N \Phi  \,+\, ....\Big]
\end{equation}

\noindent 
where $G^{MN}$ is the ten dimensional metric, ${\cal R}$ is the Ricci 
scalar curvature,  $\Phi$ is the dilaton field and 
$\kappa \sim g (\alpha^\prime)^2 $. 
We identify this ten dimensional space with an AdS$_{5}\,\times S^5$
with radius $R$ and measure (\ref{metric3}).
Furthermore we are take an AdS slice with 
"size" $z_{max}$ representing an 
infrared cut off associated with the mass of the lightest glueball.  
Note that we are also considering the dilaton to be in the $s-wave$ state, 
so we will not take 
into account variations with respect to $S^5$ coordinates. 
Thus, the action becomes 

\begin{eqnarray}
\label{Action}
S &=& {\pi^3 R^8 \over 4 \kappa^2} \int d^4x \int_0^{z_{max}} {dz\over z^3} 
 e^{-2\Phi}\nonumber\\
 &\times& \Big[
{\cal R} + ( \partial_z \Phi )^2 \,+\, \eta^{\mu\nu} \partial_\mu \Phi  
\partial_\nu \Phi\,+\, ....\Big]
\end{eqnarray}

\noindent where $\eta^{\mu\nu}$ is the four dimensional Minkowiski metric.
 
Then the momentum dependence of the bulk amplitude can be determined
using dimensional arguments. Note that the global constant 
$\, R^8 / \kappa^2\, $ associated with this action is dimensionless
and the only dimensionfull parameters are $z_{max}\,\,\sim\,1/\mu\,$ and the 
Ricci scalar $ {\cal R} \,\sim \, 1/R^2\,$. 
As $\mu\,\ll\,k$ the relevant contribution to the bulk amplitude will not involve
$z_{max}$.  Further, choosing the condition  $ 1/R \ll k $
we can disregard the term involving the Ricci scalar.
This condition does not fix completely the AdS radius $R$ and we additionally 
impose that $\mu \ll 1/R\,$. This implies that $z_{max} \gg R$.
Then, if one regularizes\cite{GKP} the divergence ($z = 0$) of the bulk action 
by cutting the axial coordinate $z$ at  $R$ , one still has a large portion of
the original AdS space: $ R \,\le \,z \,\le \,z_{max}\,$.
This guarantees that we keep the interesting AdS region which is 
an approximation for the near horizon geometry of $N$ coincident $D3$-branes,
as in the Maldacena duality.

Taking into account the normalization of the states $\vert k, u_1\rangle\,$
one sees that $\,{\cal M}_{Bulk}\,$ is dimensionally [Energy]$^{4 - n}$,  where $n$ 
is the total number of scattered particles. As $k$ is the only dimensionfull
quantity that is relevant at leading order for the bulk scattering in the 
regime considered we find:
\begin{equation}
{\cal M}_{Bulk}\,\sim\, k^{ 2 - m }\,\,.
\end{equation}

\noindent Using again the relation between bulk and boundary 
momenta (\ref{Kk}) and inserting this result in the boundary amplitude (\ref{Mb})
we get
\begin{equation}
{\cal M}_{Boundary} \,\sim \,K^{4 - \Delta } \,,
\end{equation}

\noindent where $\Delta = ( m + 2) d $ is the total scaling dimension
of the scattering particles associated with glueballs on the four 
dimensional boundary. Considering  $ K \sim \sqrt{ s} $ we find 
the expected QCD scaling behavior\cite{QCD1,BRO}
\begin{equation}
 {\cal M}_{Boundary} \,\sim \,s^{2  - \Delta/2 } 
\,.
\end{equation}

\noindent This shows that the bulk/boundary one to one mapping (\ref{ab}) , (\ref{Kk})
can be used to obtain the hard scattering behavior of high energy glueballs 
at fixed angles, from a low energy approximation of string theory. 

It is interesting to relate the different energy scales used in the above
derivation of the scattering amplitudes and check their consistency. 
The scales we discussed are 
\begin{equation}
\mu \,\,\ll \,\,{1\over R}\,\,\ll\,\, {1\over \sqrt{\alpha^\prime}}\,.
\end{equation}

\noindent Note that the relation between the AdS radius $R$ , the number of
coincident branes $N$, the string coupling constant $g$ and scale $\alpha^\prime$
is $R^4 \,\sim \, g N ( \alpha^\prime )^2$.
Then the above condition between $R$ and $\alpha^\prime$ corresponds
to the $\,$ 't Hooft limit\cite{Planar}.
Assuming that the dimensionless quantity $\mu R$ is the parameter
that relates the energy scales we find $ \sqrt{\alpha^\prime}\,=\,\mu R^2 \,$
so that the lightest glueball mass is
\begin{equation}
\mu^2 \,=\,{1\over g N \alpha^\prime}\,\,.
\end{equation}

\noindent  This result is in agreement with Maldacena and Nunez\cite{MN}. Further,  the 
above relation between the energy scales together with the condition that $k \gg 1/R\,$
and the mapping between bulk and boundary momenta (\ref{Kk}) imply that
\begin{equation}
\mu \,\,\ll K \,\ll \,{1\over R}\,\ll\,k\,\ll\,\, {1\over \sqrt{\alpha^\prime}}\,\,,
\end{equation}

\noindent so that the absolute values of $k$ are greater than those of $K$,
although the boundary scattering is a high energy process
(with respect to $\mu$) while the bulk scattering is a low  energy
process (with respect to $\,1/\sqrt{\alpha^\prime}\,$).
Furthermore  $K$ and $k$ in this regime are inversely proportional
showing a kind of infrared-ultraviolet duality as expected from  holography\cite{HOL3}.
 
\section{Glueball masses}

We can use the mapping between bulk and boundary scalar fields of the previous section 
to estimate the masses of scalar glueballs\cite{BB5}. 
In contrast to the previous case now we will not consider just one kind of glueball
field.  Rather we consider, on the boundary of the AdS slice, a sequence 
of different massive fields $\Theta_i(\vec x,t)\,$ representing the 
states of scalar glueballs with masses $\mu_i$. 
Correspondingly we have a sequence of creation-annihilation operators
$ {\bf b}_i\,\,{\bf b}_i^\dagger $ satisfying equation (\ref{canonical2})
with $ w_i(\vec K ) = \sqrt{ {\vec K}^2 + \mu_i^2}$.  

Introducing momentum operators ${\tilde \Theta}_i(K_j)\,$
with momentum $K_j$ in the interval 
${\cal E}_{j-1} \le K \le {\cal E}_j\,$ they would  not be mapped 
in a one to one relation with bulk operators ${\tilde \Phi} (k,u_p)$ 
unless $j$ is limited, since  $ i$ and $p$ are unlimited. 
Then such a mapping is possible if we introduce a restriction on the index $j$. 
The simplest choice is to take just one value for $j$.
This is obtained taking ${\cal E}_1\equiv {\cal E} $ 
large enough, which means that now $j = 1\,$. 
This recovers the one to one mapping of the previous section and in this case 
it reads
$$ {\tilde \Theta}_i( K ) \,\leftrightarrow {\tilde \Phi} (k,u_i) ,$$

\noindent where we have dropped the index of $K_1$ since this is the only relevant
boundary momentum.  

This mapping can be written explicitly in terms of bulk and boundary 
creation-annihilation  operators. We will impose the same relation
(\ref{ab}) for each pair $ {\bf a}_i \,,\,{\bf b}_i \,$.

Requiring again that equations (\ref{ab}) preserve the canonical commutation relations 
(\ref{canonical1},\ref{canonical2}) one finds that the moduli of the momenta 
are related  by (\ref{completo}) where $ 0 \le K \le {\cal E}\,$ as before.

Now we associate the size $z_{max}$ of the AdS space with the mass of the lightest
glueball which we choose to be $\mu_1$  
\begin{equation}
z_{max} \,=\, {\chi_{_{2\,,\,1}}\over \mu_1}\,,
\end{equation}
 
\noindent so that from equation (\ref{up}) we have 
\begin{equation}
u_i \,=\,{\chi_{2\,,\,i}\over \chi_{2\,,\,1}}\,\, \mu_1 .
\end{equation} 

An approximate expression for the mapping (\ref{completo}) can be obtained 
choosing appropriate energy scales.
We take ${\cal E}$ to be the string scale $ 1/ \sqrt{\alpha^\prime}$ assuming  that 
$K \ll {\cal E}$. 
On the other side we restrict 
the momenta $K$ associated with glueballs to be much larger than their masses $\mu_i$.
Then we have 
\begin{equation}
\label{region}
\mu_i \ll K \ll {1\over \sqrt{\alpha^\prime}}.
\end{equation}

\noindent In this regime the mapping (\ref{completo}) reduces to 
\begin{equation}
\label{Kk2}
k \,\,\approx\,{ u_i \,  \over 2 \,\sqrt{\alpha^\prime} K}\,. 
\end{equation}

\noindent This approximate mapping gives a high energy scaling similar to QCD\cite{BB3}.
Using the conditions (\ref{region}) together with the above mapping 
we see that the bulk momenta satisfy
\begin{equation}
u_i \ll k \ll \,\Big( {u_i \over \mu_i} \Big)\,{1\over \sqrt{\alpha^\prime}} .
\end{equation}

Note that the supergravity approximation holds for $ k \ll 1/ \sqrt{\alpha^\prime}\,$.
So in order to keep this approximation valid for all glueball operators $\Theta_i$
the factor $u_i / \mu_i $ should be nearly constant. We then impose that
 
\begin{equation}
{u_i \over \mu_i }\,=\,constant\,\,.
\end{equation}

So the glueball masses are related to the zeros of the Bessel
functions by
\begin{equation}
\label{QCD4}
{ \mu_i\over \mu_1 }\,=\,{\chi_{2\,,\,i}\over \chi_{2\,,\,1}}\,\,.
\end{equation}

Using the values of these zeros one finds the ratio of the glueball masses
for the state $0^{++}$ and its excitations. We are using the conventional notation
for these states with spin zero and positive parity and charge conjugation.
In order to compare our results 
from bulk/boundary holographic mapping with 
those coming from lattice we adopt the mass of the first state as an input.
Our results\cite{BB5} are in good agreement with lattice\cite{MASSL,MASSL2} 
and AdS-Schwarzschild black hole 
supergravity\cite{MASSG}$-$\cite{MASSG7} calculations.

It is interesting to mention that an approach to
estimate glueball masses in Yang Mills$^\ast$ from  a deformed AdS space was discussed
very recently\cite{ACEP}.

We can generalize the above results to  AdS$_{n+1}$. 
In this case massless  bulk fields are expanded in terms of the Bessel 
functions $J_{n/2} $ and the mass ratios for the $n$ dimensional "glueballs"  
are given in terms of their zeros. 
In particular for AdS$_4$, where one expects to recover results from QCD$_3$,
we find 
\begin{equation}
\label{QCD3}
{ \mu_i\over \mu_1 }\,=\,{\chi_{3/2\,,\,i}\over \chi_{3/2\,,\,1}}\,\,.
\end{equation}
\noindent Using this relation we obtain\cite{BB5} the ratio of masses
 again in good agreement with lattice\cite{MASSL,MASSL2}
 and AdS-Schwarzschild black hole
 supergravity calculations
\cite{MASSG}$^{-}$\cite{MASSG7}.

It is interesting to see if the AdS slice considered here can be 
related to the AdS-Schwarzschild black hole metric proposed by Witten\cite{Wi2}.
For the case of QCD$_3$ the metric reads 

\begin{eqnarray}
ds^2 &=& R^2 \Big( \rho^2 - {b^4\over \rho^2} \Big)^{-1} d\rho^2
+  R^2 \Big( \rho^2 - {b^4\over \rho^2} \Big) d\tau^2 \nonumber\\
&+& R^2 \rho^2 
(d\vec x)^2\,
+ \,R^2 d\Omega_5^2 \,\,,
 \end{eqnarray}

\noindent where $\rho \ge b\,$, $\,R^2 = l_s^2 \sqrt{4\pi g_s N} $ and $b $ is inversely 
proportional to the compactification radius of $S_1$ where the $\tau$ variable
is defined. 

If we qualitatively neglect the $\tau$  contribution to the 
metric in the limit of very little compactification radius
and then take the limit $\rho \gg b$ this metric is approximated by

\begin{equation}
ds^2 = { R^2 \over \rho^2}  d\rho^2
+  R^2  \rho^2 d\tau^2 
+ R^2 \rho^2 
(d\vec x)^2\,
+ \,R^2 d\Omega_5^2 \,\,.
 \end{equation}

\noindent That is an AdS$_4\times S^5\,$ space that takes a form similar
to eq. (\ref{metric}) if we
change the axial coordinate to $z\,=\,1/\rho $. 
In Witten's framework one must impose regularity conditions
at $\rho = b$ because of the presence of the horizon at this position.
In our approximation in order to retain this physical
condition we impose boundary conditions there 
and associate it to the cut of our slice ($b = 1/z_{max}$).
This AdS$_4$ slice is the one used to estimate the glueball mass ratios 
related to the three dimensional gauge theory (\ref{QCD3}). 
So we can think of our AdS$_4$  slice as a naive approximation 
to Witten's proposal.

An analogous situation could also be considered for the 
Witten's proposal to QCD$_4$. In that case the situation is more involved
because of the form of the metric coming from the compactification of 
AdS$_7\times S^4$. 

In conclusion we have seen that the bulk/boundary holographic mapping which reproduces
the high energy scaling of QCD like theories can also be applied to estimate 
glueball mass ratios. 
We hope that this mapping can be used  to describe other particle states
that may be related to some properties of QCD.

It is important to remark that one can obtain a similar result for the 
ratio of the glueball masses considering other mappings between bulk and boundary 
creation-annihilation operators instead of eq. (\ref{ab}).
For example one could take ${\bf a}_i \,= \,{\bf b}_i\,$. This would contain the solution 
$\,k = K$  implying that the masses of the glueballs are identically equal 
to the values of axial bulk momenta $u_i$. However such a trivial mapping does not 
seem to reproduce the high energy QCD scaling.

Let us mention that we used a solution for the dilaton field corresponding to
Dirichlet  boundary conditions at $z = 0 $ and $ z= z_{max}$.
This allows the existence of Bessel functions but not the divergent Neumann solutions. 
Other boundary conditions can also be considered in the same context.  
 
We have also obtained these mass ratios for scalar glueballs starting with
the same AdS slice as discussed here without using the holographic mapping 
of ref \cite{BB3} but assuming the stronger condition of relating directly the dilaton
modes with the glueball masses\cite{BB6}. The consistency between these results 
seems to indicate that the holographic mapping found before may indeed be valid
within the approximations and the energy region considered.

\section*{Acknowledgments}
The authors are partially supported by CNPq, FINEP , 
CAPES (Procad program), FAPERJ - Brazilian research agencies
and Funda\c c\~ao Jos\'e Pel\'ucio Ferreira (CCMN-UFRJ).



\end{document}